\documentclass[twocolumn,preprintnumbers,amsmath,amssymb]{revtex4}

\usepackage{graphicx}
\usepackage{dcolumn}
\usepackage{bm}

\begin{document}







\def\beq{\begin{equation}}
\def\eeq{\end{equation}}
\def\bea{\begin{eqnarray}}
\def\eea{\end{eqnarray}}
\def\ben{\begin{enumerate}}
\def\een{\end{enumerate}}
\def\la{\langle}
\def\ra{\rangle}
\def\a{\alpha}
\def\b{\beta}
\def\g{\gamma}\def\G{\Gamma}
\def\d{\delta}
\def\e{\epsilon}
\def\phi{\varphi}
\def\k{\kappa}
\def\l{\lambda}
\def\m{\mu}
\def\n{\nu}
\def\o{\omega}
\def\p{\pi}
\def\r{\rho}
\def\s{\sigma}
\def\t{\tau}
\def\L{{\cal L}}
\def\S{\Sigma }
\def\gsim{\; \raisebox{-.8ex}{$\stackrel{\textstyle >}{\sim}$}\;}
\def\lsim{\; \raisebox{-.8ex}{$\stackrel{\textstyle <}{\sim}$}\;}
\def\gtrsim{\gsim}
\def\lessim{\lsim}
\def\loc{{\rm local}}
\def\vm{v_{\rm max}}
\def\bh{\bar{h}}
\def\del{\partial}
\def\nab{\nabla}
\def\half{{\textstyle{\frac{1}{2}}}}
\def\fourth{{\textstyle{\frac{1}{4}}}}

\title{Non-equilibrium Thermodynamics of Spacetime}

\author{Christopher Eling${}^1$}
 \author{Raf Guedens${}^2$}
 \author{Ted Jacobson${}^1$}
 \affiliation{${}^1$ Department of Physics, University of Maryland\\ College Park, MD 20742-4111 USA}
  \affiliation{${}^2$ High Energy and Elementary Particle Division, University of Crete,\\
  P.O. Box 2208, GR-710 03 Heraklion, Crete, GREECE}

\begin{abstract}
It has previously been shown that the Einstein equation can be
derived from the requirement that the Clausius relation $d S =
\delta Q/T$ hold for all local acceleration horizons through each
spacetime point, where $dS$ is one quarter the horizon area change
in Planck units, and $\delta Q$ and $T$ are the energy flux across the
horizon and
Unruh temperature seen by an accelerating observer just inside the
horizon.
Here we show that a curvature correction to the entropy that is
polynomial in the Ricci scalar requires a non-equilibrium
treatment. The corresponding field equation is derived from the
entropy balance relation $dS =\delta Q/T+d_iS$, where $d_iS$ is
a bulk viscosity entropy production term that we determine by
imposing energy-momentum conservation. Entropy production can also
be included in pure Einstein theory by allowing for shear viscosity of the
horizon.
\end{abstract}

\maketitle

\section{Introduction}

A profound connection between gravitation and thermodynamics
was first suggested by the discovery in the
1970's of black hole entropy\cite{Bekenstein:1973ur}, the
four laws of classical black hole mechanics \cite{Bardeen:1973gs},
and Hawking radiation \cite{Hawking:1974sw}. But it is rather mysterious
that the Einstein equation, a hyperbolic second order
partial differential equation for the spacetime metric, has a
predisposition to thermodynamic behavior. A decade ago one of us
proposed \cite{Jacobson:1995ab} to explain this connection by reversing the
logic, using the assumed proportionality of entropy and horizon
area for all local acceleration horizons (called there local
Rindler horizons) to derive the Einstein equation as an
equilibrium equation of state. This derivation
suggests the idea that gravitation on a
macroscopic scale is a manifestation of thermodynamics of the
vacuum.

Does this thermodynamic derivation of the
Einstein equation indicate something deep,
or is it  a case of  ``assuming the answer",  or a superficial
consequence of the assumptions, or perhaps just
an accident? We
address this question here by investigating
whether and how the derivation can be
generalized to allow for the higher curvature terms
expected in the
field equation on the grounds of effective field theory\cite{Burgess:2003jk}.
One might guess that this could be done
simply by allowing for curvature to enter the ansatz for
horizon entropy. The purpose of this article is to investigate
whether
this is so.
We find that entropy dependence on the Ricci scalar can indeed be
accommodated, but it requires a change of setting from equilibrium
to non-equilibrium thermodynamics.

We begin by reviewing the two hypotheses on which the derivation of
the Einstein equation of state given in Ref.~\cite{Jacobson:1995ab} is based.
Next we present that derivation,
and then generalize it to allow for
dependence of the entropy on the Ricci scalar. We end with
several comments.

The motivating idea is that the origin of the thermodynamic behavior
of black holes is to be traced to the thermal nature of the
Minkowski vacuum. The vacuum is the ground state of the
generator of time translations, but it is thermal with respect to any
generator of Lorentz boosts. More precisely, restricted to the
``Rindler wedge" $x>|t|$ (using Minkowski coordinates) the vacuum
density matrix for a relativistic quantum field has the form of the
canonical ensemble (Gibbs state) $\rho=Z^{-1}\exp(-H_B/T)$,
where $H_B$ is the boost hamiltonian and the ``temperature" is
$T=\hbar/2\pi$.~\cite{Bisognano:1975ih, Unruh:1983ac}.
This
``temperature" does not have dimensions of energy, because the boost
$H_B$  generates translations of a dimensionless hyperbolic
angle, rather than of a time. When re-scaled to generate
proper time translations on
the worldline of a uniformly accelerated observer
this becomes the Unruh temperature
$T_U=\hbar a/2\pi$~\cite{Unruh:1983ac} , where $a$ is the acceleration.

The past boundary $x=t<0$ of the Rindler wedge is a lightlike
hyperplane in spacetime, forming a causal boundary or horizon for the past of
the ``bifurcation plane" $x=t=0$.
The state behind the horizon is hidden to outside observers with
access only to a spatial slice bounded by this plane. For such
observers the relevant state is the density matrix $\rho$. The
entropy of $\rho$ is infinite in standard quantum field theory, but
when UV regulated it is proportional to the area of the bifurcation
plane and depends on the number and nature of the quantum
fields~\cite{Bombelli:1986rw}. The similarity to Bekenstein-Hawking
black hole entropy, which is universally given by the horizon area
divided by $4\hbar G$, motivates our first hypothesis: we suppose
that in quantum spacetime---whatever that is---there is a universal
entropy density $\alpha$ per unit horizon area. We imagine that
$\alpha$ indeed depends on the number and nature of quantum fields,
if any such freedom exists in the underlying fundamental theory.

Our second hypothesis concerns the relation between this entropy and
the flux of boost energy across the Rindler horizon. Under a small
perturbation of any Gibbs state at temperature $T$, the variation of
the entropy is related to the variation of the mean energy by $\d
S=\d\la E\ra/T$. When the energy is ``heat" this identity expresses
the Clausius relation $dS=\d Q/T$. The detailed nature of energy
that flows into a Rindler wedge cannot be examined by the outside
observers, hence it can be considered by them as heat that has
flowed into the thermal system behind the horizon. This motivates
our second hypothesis: the Clausius relation holds for all local
causal horizons (defined precisely below) in a sufficiently small
neighborhood of the bifurcation plane, with $\d Q$ interpreted as
the mean flux of boost energy across the horizon and $dS$ as the
difference of area for the wedge with and without the boost energy
flux.
This hypothesis is inconsistent with
the assumption of a fixed, flat spacetime,
since a Rindler horizon has fixed area,
but it is consistent with the horizon-focusing
effects of spacetime curvature
provided the Einstein equation holds~\cite{Jacobson:1995ab}.

We define a local causal horizon at a point $p$ as follows: choose a
spacelike 2-surface patch $B$ including $p$, and choose one side of
the boundary of the past of $B$. Near $p$ this boundary is a
congruence of null geodesics orthogonal to $B$. These comprise the
horizon.
We choose $B$ such that this congruence has
vanishing expansion $\theta$ and shear $\sigma_{ab}$ at $p$, which
corresponds to an equilibrium state at $p$.
We consider
transitions that terminate in this equilibrium state.

To define the heat flux and temperature we employ an approximate
boost Killing vector field $\chi^a$ that vanishes at $p$ and whose
flow leaves invariant the tangent plane $B_p$ to $B$ at $p$. That
is, the covariant derivative $\chi_{a;b}$ is a timelike
antisymmetric tensor orthogonal to $B_p$ at $p$. We normalize
$\chi^a$ by $\chi _{a;b}\chi ^{a;b}=-2$, like the usual  boost
Killing vector $x\partial_t+t\partial_x$ in Minkowski spacetime.
These conditions at $p$ would determine a unique boost Killing
vector up to sign in flat spacetime. In a generic curved spacetime
there are no Killing vectors, so the most we can do is solve
Killing's equation $ \chi_{a;b} + \chi_{b;a}=0$ with this ``initial
data"  out to some order in the neighborhood of $p$.  In Riemann
normal coordinates $\{x^\a\}$ based at $p$ these initial conditions
determine the zeroth and first order parts of $\chi^a$, while
Killing's equation imposes that the second order part vanishes.
Generally the equation cannot be satisfied at third order.

We choose the direction of $\chi^a$  to be future pointing on the
causal horizon. In flat spacetime, it would then be related on the
horizon generator through $p$ to the affinely parametrized horizon
tangent vector $k^a$ via $\chi^a = -\lambda k^a$, where $\lambda$
is an affine parameter that is negative and increasing along the
horizon and vanishes at $p$. Up to the $O(x^3)$ ambiguities in
$\chi^a$, the same is true in the curved spacetime.

As motivated above, we define the heat as the mean flux of the boost
energy current of matter across the horizon, \beq \d Q = \int
T_{ab}\chi ^a d\Sigma^b, \eeq where $T_{ab}$ is the expectation
value of the matter stress tensor. (We omit the bracket notation for
brevity.) This and all subsequent integrals are taken over a short
segment of a thin pencil of horizon generators centered on the one
that terminates at $p$. Using the relation $\chi ^a = -\lambda k^a$
(which holds on the generator through $p$ up to order $O(x^3)$) and
$T=\hbar/2\pi$ we thus have \beq \frac{\d Q}{T} = (2\pi/\hbar)\int
T_{ab}k^a k^b (-\lambda)d\lambda d^2A. \label{dQ/T} \eeq

The entropy change $\d S=\alpha\, \delta A$ is determined by the
area change of the horizon, \beq \d A = \int \theta \, d\lambda
d^2A, \label{dA} \eeq where $\theta=d(\ln d^2A)/d\lambda$ is the
expansion of the congruence of null geodesics generating the
horizon. Using the Raychaudhuri equation
\beq \frac{d\theta}{d\lambda} =
-\frac{1}{2}\theta^2-\sigma_{ab}\sigma^{ab}-R_{ab}k^a k^b
\label{Ray}\eeq
and the assumed vanishing of $\theta$ and $\sigma_{ab}$ at $p$ we
have \beq \theta= -\lambda R_{ab}k^a k^b + O(\lambda^2).
\label{theta} \eeq The entropy change is thus given to lowest order
in $\lambda$ by \beq \d S = \alpha \int R_{ab}k^a k^b
(-\lambda)d\lambda d^2A. \label{dS}\eeq

If we now require that $\d S = \d Q/T$ hold for all local Rindler
horizons through all points $p$, we infer that the integrands of
(\ref{dQ/T}) and (\ref{dS}) must match for all null vectors $k^a$.
The integrands are both first order in $\lambda$, and equality of
the coefficients of $\lambda$ implies the relation
\beq R_{ab}  +
\Phi g_{ab} = (2\pi/\hbar\alpha)T_{ab} \label{eos}
\eeq
where $\Phi$
is a so far undetermined function. To determine $\Phi$ we require
that the matter stress tensor is divergence free, corresponding to
the usual local conservation of matter energy. Taking the divergence
of both sides of (\ref{eos}) and using the contracted Bianchi
identity $R_{ab}{}^{;a}=\frac{1}{2} R_{,b}$ we then find that
$\Phi=-\frac{1}{2}R - \Lambda$, where $\Lambda$ is a constant.
Therefore (\ref{eos}) corresponds to the Einstein equation
$R_{ab}-\frac{1}{2} Rg_{ab}-\Lambda g_{ab}=8\pi G T_{ab}$ with
(undetermined) cosmological constant $\Lambda$ and with Newton's
constant determined by the universal entropy density $\alpha$,
\beq G=(4\hbar\alpha)^{-1}. \eeq
Note that the entropy area density is thus
{\it universally}
$(4\hbar G)^{-1}$, no matter what is the nature and number of
quantum fields, in agreement with the Bekenstein-Hawking black hole
entropy. This completes our review of Ref.~\cite{Jacobson:1995ab}.

We now seek the thermodynamic equation of state if the entropy
density is taken to be $\alpha$ times a function $f(R)=1+O(R)$ of
the Ricci scalar,
instead of a constant $\alpha$ as before. In this case the entropy
change is given in analogy with (\ref{dA}) by \beq \d S = \alpha
\int (\theta f+ \dot{f}) \, d\lambda d^2A, \label{dSf} \eeq where
the overdot signifies derivative with respect to $\lambda$. If the
expansion $\theta$ vanishes at $p$ then the integrand of (\ref{dSf})
at $p$ is $\dot{f}=f'(R)k^a R_{,a}$, which is generally non-zero.
This cannot match the $\d Q/T$ integrand which is of order
$\lambda$. Thus the Clausius relation implies that $\theta(p)$ must
be non-vanishing, so as to cancel off the derivative of
$f$\footnote{This point was missed in Ref.~\cite{Jacobson:1995ab},
where it was therefore incorrectly asserted that the derivation of
the equation of state goes through without modification.}. That is,
we must have
\beq (\theta f + \dot{f})(p)=0. \label{equil} \eeq
This means that the causal horizon must be defined as the boundary of the
past of a 2-surface $B$ that is warped at $p$ such that
(\ref{equil}) is satisfied.
This does not coincide with the approximate Killing horizon as closely
as when the expansion vanishes.
Nevertheless, the approximate Killing vector is still
related to the tangent of the horizon generator through $p$ by
$\chi^a = -\lambda k^a$, up to the $O(x^3)$ ambiguity of
$\chi^a$.

Since the area of the horizon is changing at $p$, it seems at
first that the ``system" does not approach an equilibrium state at
$p$, even though the {\it entropy} is instantaneously stationary
there. However, the relevant notion of time here is the Killing
flow.
The relation between the affine parameter $\lambda$ and
Killing parameter $v$  on a Rindler horizon
is $\lambda=-\exp(-v)$, so the point $p$
occurs at infinite Killing time. Even if $\theta(p)\ne0$, the rate
of change of area with respect to Killing time vanishes like
$\sim\exp(-v)$ as $p$ is approached. This can be considered an
approach to equilibrium. In the special case  $\theta(p)=0$
considered previously, the
expansion vanishes at twice this rate, i.e. as
$\sim\exp(-2v)$. The slower decay rate in the general case
suggests that equilibrium thermodynamics may not apply, so that
the Clausius relation may not hold. Instead we may have $dS>\delta
Q/T$, or more precisely the {\it entropy balance relation}
\beq dS=\delta Q/T+d_iS,\label{Sbal} \eeq
where $d_iS$ is the entropy developed internally in the system as
a result of being out of equilibrium~\cite{deGroot}.
The entropy production rate
vanishes at $p$, since that is an equilibrium point, so we expect
the rate is of order $\lambda$.


To extract the $O(\lambda)$ term in the integrand of (\ref{dSf}) we
differentiate with respect to $\lambda$ and use (\ref{equil}),
\beq {\textstyle\frac{d}{d\l}}(\theta f + \dot{f})|_{\lambda=0} =
\dot{\theta}f -f^{-1}\dot{f}^2 + \ddot{f}. \eeq
Using the Raychaudhuri equation (\ref{Ray}) and the geodesic
equation $k^a k^b{}_{;a}=0$  this takes the form
\beq
 -k^ak^b(fR_{ab} - f_{;ab} +f^{-1} f_{,a}\, f_{,b})-\half
f\theta^2.\label{S1} \eeq
Were there no entropy production $d_iS$, the Clausius relation would
imply that (\ref{S1}) must be equal to the coefficient of $\lambda$
in the heat flux integrand of (\ref{dQ/T}) for all null vectors
$k^a$. It would follow that in place of (\ref{eos}) we have
\beq fR_{ab} - f_{;ab} +\textstyle{\frac{3}{2}}f^{-1} f_{,a}\,
f_{,b} + \Psi g_{ab} = (2\pi/\hbar\alpha)T_{ab}, \label{eosf} \eeq
where (\ref{equil}) has
been used to re-express the $\theta^2$ term
and  $\Psi$ is a so far undetermined function.
We now show that this is inconsistent with energy conservation.

We
require, as before, that the matter stress tensor is divergence
free, so the divergence of the left hand side of (\ref{eosf}) must
vanish. Using the contracted Bianchi identity, the commutator of
covariant derivatives $2v^c{}_{;[ab]}
= R_{abd}{}^c v^d$, and defining
$\cal L$ by $f=d{\cal
L}/dR$, we find
\beq(fR_{ab} - f_{;ab})^{;a} =
(\frac{1}{2}{\cal L} -\Box f)_{,b}.
\eeq
Thus we must have \beq \Psi= \Box f-\frac{1}{2}{\cal L} - \Theta,
\label{Psi} \eeq where the gradient of $\Theta$ matches the divergence of
the remaining term in (\ref{eosf}),
\beq
\Theta_{,b}=
(\frac{3}{2f}f_{,a}f_{,b})^{;a}.
\label{Theta}
\eeq
This reveals a contradiction, however, since the right hand
side of (\ref{Theta}) is generally not the gradient of a scalar.

We propose that this contradiction with energy-momentum conservation
is resolved by the entropy production term $d_iS$ in (\ref{Sbal}).
Examination of (\ref{S1}) shows that the problematic term would be
canceled if we set \beq d_iS=\int\s\, d\l d^2A \eeq with entropy
production density
\beq \s=-\textstyle\frac{3}{2}\alpha f^{-1}\dot{f}^2\,
\l=-\textstyle\frac{3}{2}\alpha f\theta^2\, \l \label{Sprod} \eeq
(using (\ref{equil})).
In terms of
the expansion with respect to Killing parameter,
$\tilde{\theta}=\theta(d\lambda/dv)$, we have
 \beq
\s\,  d\lambda=\textstyle\frac{3}{2}\alpha f\tilde{\theta}^2\, dv.
\label{Sprodv} \eeq
This is just like the entropy production term for a fluid at
temperature $T$ due to a bulk viscosity $\eta=(3/2)\alpha
fT$~\cite{deGroot}. Putting $T$ equal to the boost temperature
$\hbar/2\pi$ yields $\eta=3\hbar\alpha f/4\pi$. With this for
$d_iS$, equation (\ref{Sbal}) at $O(\l)$ implies the equation of
state
\beq fR_{ab}- f_{;ab} +(\Box f-\frac{1}{2}{\cal L})g_{ab}=
(2\pi/\hbar\alpha) T_{ab}. \label{Leom} \eeq
This coincides with the equation of motion arising from the
Lagrangian $(\hbar\alpha/4\pi){\cal L}(R)$ for which the entropy
density of a stationary black hole horizon is $\alpha
f(R)$~\cite{6}. The thermodynamic equation of state is therefore
again consistent with the Lagrangian field equation, as in the pure
GR case. We conclude with a number of remarks.

1. Given an entropy functional of the macroscopic variables of an
ordinary thermodynamic system, one can normally derive the equation
of state from the Clausius relation together with the first law of
thermodynamics. In our case the first law was not explicitly
invoked. However, to fix the trace part of the field equation we
required that the energy-momentum tensor of matter is divergence
free, which expresses the local conservation of matter energy and is
hence a form of the first law.

2. It is common in near-equilibrium thermodynamics to deduce the
general form of entropy production terms quadratic in the gradients
of state variables, but the coefficients of those terms are
phenomenological and depend on details of the
microphysics~\cite{deGroot}. It may therefore seem puzzling that
here we deduce also the bulk viscosity $\eta$. However, $\eta$ is
precisely $3\hbar/4\pi$ times the entropy density $\alpha f$, and
$\alpha$ is purely phenomenological in our derivation. The only
puzzle is therefore the simple relation between entropy density and
bulk viscosity.
Presumably general covariance lies at the root of this.

3. In pure GR the bulk viscosity appears to become
$3\hbar\alpha/4\pi$, but this is not correct since $\theta(p)=0$,
so there is no $O(\lambda)$ entropy production term in the integrand
of (\ref{dSf}). By contrast, for  globally defined {\it black hole}
horizons it has been shown~\cite{Price:1986yy} that the bulk
viscosity is {\it negative} and equal to $-1/16\pi
G=-\hbar\alpha/4\pi$.

4. In the presence of curvature contributions to the entropy, we had
no choice but to use a non-equilibrium entropy balance relation, due
to the nonzero expansion at $p$ (\ref{equil}). Could a
non-equilibrium description be used voluntarily also for pure GR,
where the entropy is just the area? We cannot allow for expansion at
$p$, since without the derivative of $f$ to balance the expansion,
there would be a first order term in the entropy change $\delta S$
not matched by the heat flux term. But how about allowing for shear
at $p$? We initially set the shear to zero on the grounds that it
was required by equilibrium at $p$. This is not valid however, since
the shear defined with respect to Killing parameter would in any
case vanish. The consequence of nonzero shear at $p$ is to introduce
into (\ref{theta}) a term of the form $-\lambda
\sigma_{ab}\sigma^{ab}$.
This additional term can be written in terms of derivatives of
$k^a$, which can be independently chosen at $p$.
The $k^a$ parts of the Clausius relation imply  the Einstein equation as
above, but the $\partial k^a$ part is satisfied only if the shear vanishes.
However,
nonzero shear at $p$ is allowed if we include an internal entropy
production  term $\alpha \sigma_{ab}\sigma^{ab}$
corresponding~\cite{deGroot} at temperature $T(=\hbar/2\pi)$ to a
shear viscosity $T\alpha/2=\hbar\alpha/4\pi=1/16\pi G$,
just as for a black hole
horizon~\cite{Price:1986yy}.

5. Under what conditions are the higher curvature terms meaningful
in the thermodynamic interpretation? Let the entropy expansion
coefficients $\beta_i$ be defined by $f(R)= 1 + \beta_1 R +\beta_2
R^2 + \cdots$,  and let the curvature length scale at $p$ be set by
$L_c$. Then $f_{,a}f_{,b}$ in (\ref{eosf}) is $O(\beta_1/L_c^2)$
smaller than $f_{,ab}$ and the $O(R^2)$ part of $fR_{ab}$,  which
are smaller than the Einstein term by the same factor. Our
assumption that the entropy balance relation holds over an arbitrary
small patch of local causal horizon is questionable due to quantum
fluctuations at the Planck scale. If we assume only that the
relation holds over a region of some minimum coordinate size $\e$ in
Riemann normal coordinates, the definition of heat flux will be
fuzzy. How fuzzy? The Killing vector is ambiguous at $O(x^3)$, while
the lowest order part is $O(x)$.
The corresponding fuzziness in the
heat flux is therefore of relative size $O(\e^2/L_c^2)$.
The ratio of this to the relative suppression of the $L_c^{-4}$
and $L_c^{-6}$  terms in the field equation is
$\e^2/\beta_1$ and $(\e L_c/\beta_1)^2$
respectively.
If $\beta_1$ is sufficiently large ($\beta_1\gg \e L_c$)
all the curvature corrections
will be meaningful as long as $\e\le L_c$.
This is fine for a ``theoretical laboratory,"
however, dimensional analysis suggests that in nature we have
$\beta_1\sim\e^2\sim L_{\rm Planck}^2$. If this is the case, then the
fuzz is always comparable to the $L_c^{-4}$ terms
and much larger then the $L_c^{-6}$ ones.
This conclusion could be evaded if it were somehow correct to
define the heat flux using the well-defined horizon generating
vector $-\lambda k^a$ rather than the fuzzy Killing vector.

6. Having looked at the simplest case it would be interesting to
allow for other types of curvature contributions that can arise. One
might imagine that results similar to those found here would apply.
However,  since the entropy can involve not just the curvature
scalar but different projections of curvature into the horizon or
normal to it, this is certainly not clear.

\section*{Acknowledgments}
We would like to thank J.R.~Dorfman, S. Hayward, and A. Roura for
helpful discussions.
This research was supported in part by the NSF under grant
PHY-0300710.

\end{document}